# Magnetic field induced splitting and polarization of monolayer-based valley exciton-polaritons

Nils Lundt[1], Evgeny Sedov[2,3], Max Waldherr[1], Martin Klaas[1], Heiko Knopf[4,5], Mark Blei[6], Sefaating Tongay[6], Sebastian Klembt[1], Takashi Taniguchi[7], Kenji Watanabe[7], Ulrike Schulz[4], Alexey Kavokin[8,9,2], Sven Höfling[1,10], Falk Eilenberger[4,5,11], Christian Schneider[1]

[1]*Technische Physik and Wilhelm-Conrad-Röntgen-Research Center for Complex Material Systems, Universität Würzburg, D-97074 Würzburg, Am Hubland, Germany*

[2]*School of Physics and Astronomy, University of Southampton, Highfield, Southampton, SO171BJ, UK*

[3]*Vladimir State University named after A.G. and N.G. Stoletovs, Gorky str. 87, 600000, Vladimir, Russia*

[4]*Fraunhofer Institute of Applied Optics and Precision Engineering IOF, Center of Excellence in Photonics, Albert-Einstein-Straße 7, 07745 Jena, Germany*

[5]*Institute of Applied Physics, Abbe Center of Photonics, Friedrich Schiller University, Albert-Einstein-Straße 15, 07745 Jena, Germany*

[6]*School for Engineering of Matter, Transport, and Energy, Arizona State University, Tempe, Arizona 85287, USA*

[7]*National Institute for Materials Science, Tsukuba, Ibaraki 305-0044, Japan*

[8]*Westlake University, 18 Shilongshan Road, Hangzhou 310024, Zhejiang Province, China*

[9]*Institute of Natural Sciences, Westlake Institute for Advanced Study, 18 Shilongshan Road, Hangzhou 310024, Zhejiang Province, China*

[10]*SUPA, School of Physics and Astronomy, University of St. Andrews, St. Andrews KY 16 9SS, United Kingdom*

[11]*Max Planck School of Photonics, Germany*

**Atomically thin crystals of transition metal dichalcogenides are ideally suited to study the interplay of light-matter coupling, polarization and magnetic field effects. In this work, we investiagte the formation of exciton-polaritons in a MoSe$_2$ monolayer, which is integrated in a fully-grown, monolithic microcavity. Due to the narrow linewidth of the polaritonic resonances, we are able to directly investigate the emerging valley Zeeman splitting of the hybrid light-matter resonances in the presence of a magnetic field. At a detuning of -54.5 meV (13.5 % matter constituent of the lower polariton branch), we find a Zeeman splitting of the lower polariton branch of 0.36 meV, which can**

**be directly associated with an excitonic g factor of 3.94±0.13. Remarkably, we find that a magnetic field of 6T is sufficient to induce a notable valley polarization of 15 % in our polariton system, which approaches 30% at 9T. Strikingly, this circular polarization degree of the polariton (ground) state exceeds the polarization of the exciton reservoir for equal magnetic field magnitudes by approximately 50%, as a consequence of enhanced relaxation of bosons in our monolayer-based system.**

Atomically thin transition metal dichalcogenides are a new exciting class of materials, which are uniquely suited to study advanced phenomena related to light-matter coupling. Their huge oscillator strength and likewise large exciton binding energies allow for the observation of strong coupling phenomena in microcavities with a single monolayer[1], even up to room temperature[2-4]. The materials furthermore comprise the so-called valley degree of freedom, arising from their unique band structure. Due to the combination of strong spin-orbit coupling and inversion symmetry breaking, exciton spin orientations are inverted at opposite K points at the corners of the hexagonal Brillouin zone[5,6] and the K and K' valleys can be selectively addressed by σ+ and σ- circular polarized light[7,8]. The selective polarization of the valleys has been demonstrated by various resonant- and non-resonant optical techniques[9–12], and it has been shown that spin-relaxation can be suppressed in the strong light-matter coupling regime[13–17].

In the presence of external magnetic fields, a valley Zeeman splitting of exciton and trion resonances in transition metal dichalcogenide (TMDC) monolayers has been identified[18–20], similar to excitons in conventional semiconductors with two spin-projections. This splitting, which lifts the energy degeneracy between K and K' excitons, arises from a complex interplay between carrier spin and orbital momentum, which is influenced by external magnetic fields. This effect has been used to manipulate the valley polarization in TMDC monolayers[21,22]. While the effect is well investigated in pure exciton resonances, it has not been observed for exciton-polaritons, which are promising model systems to investigate bosonic condensation phenomena and exploit the spin-valley properties of TMDC monolayers in coherent macroscopic quantum states.

Here, we study the magnetic behavior of a strongly coupled $MoSe_2$ monolayer exciton in a high-Q monolithic microcavity. When applying high magnetic fields up to 9 T, we observe that the characteristic valley Zeeman splitting of the K and K' valley excitons is preserved and transferred to the exciton-polariton modes. The splitting is renormalized by the presence of the cavity photon, but nevertheless, it can still be clearly observed in polarization resolved measurements due to the narrow polariton linewidth. Lifting the valley degeneracy leads to a population imbalance in the two valley Zeeman-split resonances, which yields a significant degree of circular polarization of approximately 30% at 9 T in the ground state, which exceeds the polarization of the bare exciton under comparable conditions. This strongly indicates polarization selective, enhanced scattering driven by final state stimulation.

**Sample Structure and characterization**

The studied sample structure is schematically depicted in Fig. 1a. The microcavity is built by transferring a MoSe$_2$ monolayer with a dry-gel method[23] onto an SiO$_2$/TiO$_2$ bottom distributed Bragg reflector (DBR), grown onto a quartz glass substrate by physical vapor deposition. The bottom DBR consists of 10 pairs and has a stop band center at 750 nm. The monolayer was capped with a mechanically exfoliated flake of hexagonal boron nitride (h-BN, about 10 nm thick), to protect the monolayer from the subsequent overgrowth processing conditions. The top DBR (9 pairs) was grown by plasma-assisted evaporation (PAE) at mild processing temperatures of 80° C. The complete, but empty microcavity exhibits a clear, parabolic dispersion relation with a ground mode at 1.66 eV as shown in the in-plane momentum-resolved reflectivity measurement (Fourier plane imaging method) in Fig 1b. The line spectrum at zero in-plane momentum (Fig. 1c) reveals a resonance linewidth Δλ of 0.151 nm as presented in figure 1c. This is equivalent to a Q-factor of 4960 ($Q = \frac{\lambda}{\Delta\lambda}$). Fig. 1d shows a microscopy image of the final sample with the monolayer of MoSe$_2$ and hBN capping adjacent to bulk remnants taken before the growth of the top DBR.

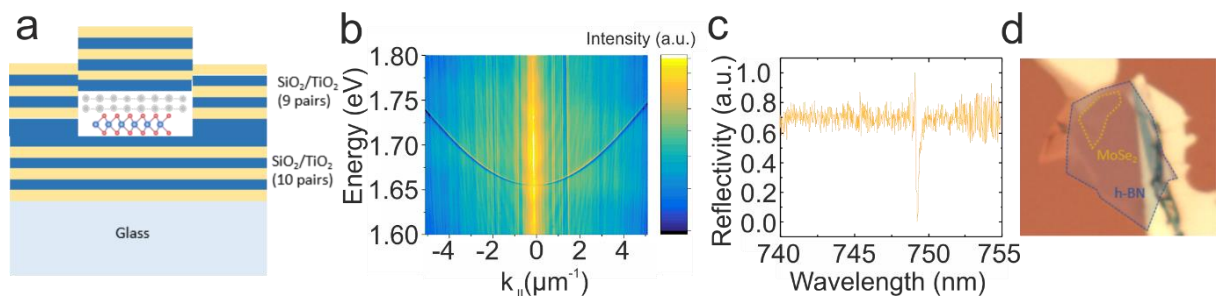

*Fig. 1 | **Sample Structure.** a) Schematic of the investigated microcavity structure, which consists of a MoSe$_2$ flake between SiO$_2$/TiO$_2$ mirrors. b) Reflectivity spectrum showing the photon mode. c) Line spectrum at zero in-plane momentum revealing the high Q-factor of approx. 5000. d) Microscopy image of the monolayer with h-BN capping in the fully-grown microcavity.*

**Bare monolayer and photon mode in a magnetic field**

First, we investigate the polariton constituents separately in a magnetic field via photoluminescence spectroscopy. We excite with a cw laser at 532 nm. The experiments were performed at 5 Kelvin. Figure 2a depicts the results of a bare monolayer. The energy positions have been extracted from a Lorentz fit of the exciton emission, which has been resolved in the two circular polarization directions with a standard λ/4 wave plate configuration. We obtain a splitting of 1.8 meV at 9T, which corresponds to an excitonic g factor of 3.58 ± 0.1. Values in the literature range from 3.1 to 4.3, depending on flake environment[17-19]. The degree of circular polarization (DOCP) is calculated as $(I_+ - I_-)/(I_+ + I_-)$. Its dependence is shown in Fig. 2b. The DOCP originates in the energy splitting, and monotonically increases up to approximately 20 % at 9 T. The phenomenological behavior can be captured by a semiclassical Boltzmann model, which is introduced in the theory section of the manuscript. We furthermore probed the reaction of a pure photonic resonance from a spatial position close to the monolayer in figure 2c. As expected for a photonic system, no systematic energy shift nor a splitting can be observed.

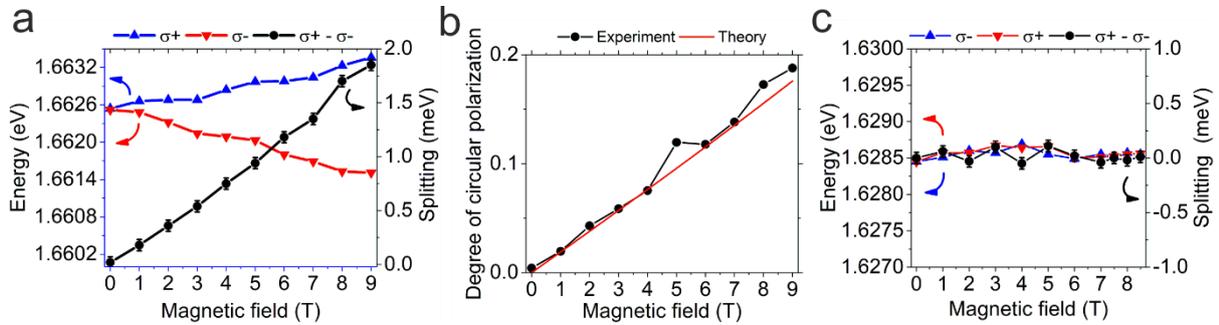

*Fig. 2 | **Magnetic field dependence of exciton and photon.** a) Polarization resolved energy positions (extracted from a Lorentz fit of the photoluminescence spectrum) from bare monolayer excitons. A clear Zeeman splitting can be observed. Error bars from spectrometer resolution. b) Degree of circular polarization with increasing magnetic field measured (black) and simulated (red, from equations 1 and 2). c) Emission energy and polarization splitting of a photon mode in the vicinity of the investigated monolayer.*

**TMDC based exciton-polaritons in a magnetic field**

Now we turn to a study of the microcavity including the MoSe$_2$ monolayer by in-plane momentum-resolved photoluminescence. Figure 3a shows a photoluminescence (PL) spectrum of the completed microcavity sample, where we plot the emission energy as a function of the in-plane momentum. The cavity mode outside the monolayer region at 1.66 eV is faintly visible. Due to the finite thickness of the h-BN flake, the uncoupled cavity resonance was shifted to 1.621 eV (also visible in the spectrum). Additionally, at the spot of the monolayer, the photon mode is changed even further due to a change in the effective cavity length via the refractive index of the monolayer, which we calculated via transfer matrix method to be 1.608 eV. In addition, various new and discrete modes have appeared below the uncoupled exciton and cavity resonances, which we associate with the formation of spatially confined exciton-polaritons by finite lateral size effects. The fragmentation of the spectrum is a strong indication of photonic disorder in our system. The monolayer extension locally confines the optical mode in our structure[24]. From the mode splitting, we can estimate an effective trapping length of ~ 6.4 μm.

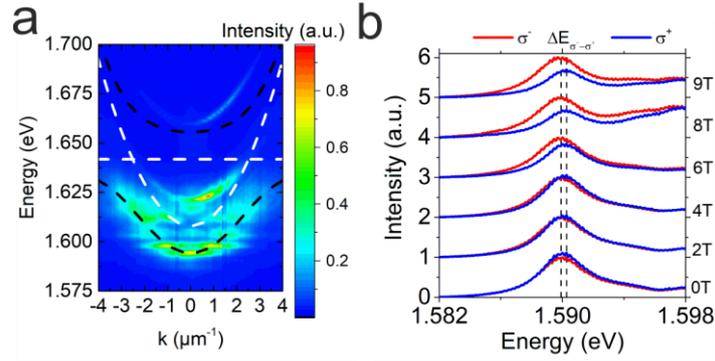

*Fig. 3 | **Polariton dispersion and emission changes in a magnetic field.** a) Angle-resolved characterization of the polariton modes. White dashed lines correspond to exciton and photon resonances respectively. Black dashed lines are the result for the polariton modes from a coupled oscillator model. b) Energy resolved polaritonic ground state emission intensity magnetic field series from 0 T to 9 T (each spectrum normalized to the sigma minus one).*

In order to model the polariton modes of the system, we applied a standard coupled oscillator model[25] and use the uncoupled exciton (1.6626 eV) as well as the cavity resonance determined from the off-flake position (1.608 eV) and the measured lower polariton branch (1.5943 eV) as input parameters. The vacuum Rabi splitting can then be extracted to be 51.1 meV, with an exciton-photon detuning of -54.5 meV. The Hopfield coefficient determines the exciton fraction in the polariton state and its value is $|X|^2$ = 0.1354.

**Polariton valley Zeeman splitting**

The exciton-polariton inherits its properties from both the exciton and photon mode, and should exhibit a notable effect in an external magnetic field: Thus, we investigate our completed microcavity sample in the presence of an applied magnetic field up to 9 T in the Faraday configuration. Fig. 3b and c show a comparison of the polarization resolved ground state spectra, recorded at various magnetic fields. While we normalized all spectra in Fig 3b to visualize the emergent splitting in energy, in Fig 3c the normalization was omitted to highlight the induced circular polarization. A clear Zeeman splitting emerges and sizeable circular polarization buildup can be observed. We emphasize, that both effects are in striking difference to the pure photonic case exhibited in Fig. 2c, and that the Zeeman-split emission can be attributed to the effect of light-matter hybridization.

The results of this systematic series is summarized in Fig. 4a. The Zeeman splitting increases linearly with the applied magnetic field up to a value of approximately. 0.36meV. In order to quantify the strength of the Zeeman splitting, we approximate our data by the equation $\Delta E = |X|^2 \Delta E_R$, where $\Delta E_R = g\mu_B B$ is the splitting strength of the reservoir excitons; $g$ denotes the excitonic g-factor, and $|X|^2$ is the exciton fraction of our polaritons. Since the diamagnetic shift of excitons in TMDC monolayer for magnetic fields <10 T is negligible, we can safely neglect the magnetic field dependence of the $|X|^2$ in our analysis. Fitting our data yields an exciton g factor of 3.94±0.13, which is in line with the characterization in Fig. 2 on bare excitons and the existing literature[17-19], considering the range of measured values due to effects of adjacent layers on the effective g factor.

The splitting of the exciton, as well as the polariton resonances intrinsically yields an unbalance of pseudospins (and thus polarization) in the exciton- and polariton reservoir. This is well reflected in the emergence of circular polarization in our system at increased magnetic fields. In Fig. 4b, we plot the degree of circular polarization for the polariton ground state (integrated over the full mode). Despite the reduced Zeeman-splitting of this mode, the polarization is significantly more pronounced as for the case of the bare exciton in Fig 2b. Moreover, as depicted in Fig 4c, the degree of circular polarization for our polariton modes is always the highest in the ground state (at k=0) for every applied magnetic field, and can even exceed the polarization of excited polariton states and the exciton reservoir by a factor of two.

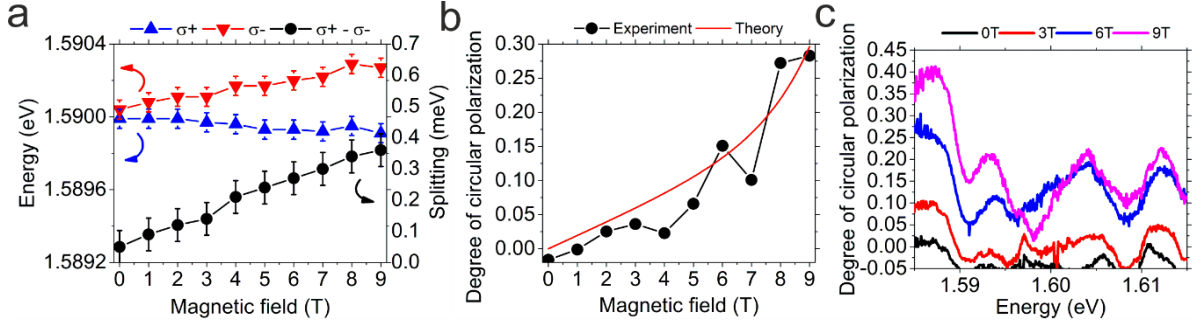

Fig. 4 | *Magnetic field series and valley polarization buildup.* a) Polarization resolved energy position of the polariton ground state depending on an external magnetic field. The induced Zeeman splitting rises monotonically to approximately 0.36 meV. b) Circular polarization buildup with increased magnetic field measured (black) and simulated (red) with respect to Eqs. (1)–(2). c) DOCP along the magnetic field series, evaluated for each energy. The groundstate always shows significantly increased polarization.

Quantitatively, the DOCP can be understood as a consequence of spin-valley relaxation in the exciton reservoir, as well as in the polariton modes. We confront our data with this hypothesis, and provide a semi-analytical model to our data based on a set of the coupled semiclassical Boltzmann equations for the occupations of the polariton condensate and exciton reservoir modes:

$$\frac{dN^\pm}{dt} = W N_R^\pm (N^\pm + 1) - \frac{N^\pm}{\tau} \mp \left( \frac{N^+}{\tau_{s+}} - \frac{N^-}{\tau_{s-}} \right), \tag{1}$$

$$\frac{dN_R^\pm}{dt} = P_\pm - W N_R^\pm (N^\pm + 1) - \frac{N_R^\pm}{\tau_R} \mp \left( \frac{N_R^+}{\tau_{R,s+}} - \frac{N_R^-}{\tau_{R,s-}} \right). \tag{2}$$

In Eqs. (1)–(2), $N^\pm$ and $N_R^\pm$ are the occupations of the polariton modes at k=0 and the occupations of the high $k$ reservoir in the two spin components, respectively. $W$ is the stimulated scattering rate from the reservoir to the polariton ground state. $\tau$, $\tau_R$ are the lifetimes of the ground state polaritons and the reservoir excitons. To take into account the effect of the magnetic field, we introduce the B-dependent spin-relaxation times for polaritons and excitons as following: $\tau_{s\pm}(B) = \tau_s \exp[\pm \Delta E(B)/k_B T]$ and $\tau_{R,s\pm}(B) = \tau_{R,s} \exp[\pm \Delta E_R(B)/k_B T]$, where $\tau_s$ and $\tau_{R,s}$ correspond to the relaxation times in the absence of the magnetic field. $P_\pm$ are the pumping rates of the reservoir. We further assume that both reservoirs are equally efficiently populated by the linear pump $P_\pm = P$. The DOCP is finally calculated by evaluating the imbalance of quasi-particle densities, both in the reservoir and the polariton ground state as a function of the pumping rate. The results of this modelling are plotted with the experimental data in Fig. 2b for the bare monolayer excitons and 4b for the polaritons in the lowest energy state. Values of the parameters used for modelling are given in Ref. 26. Strikingly, we find that the model captures the increasing circular polarization in the presence of the magnetic field, both for the free exciton (Fig 2b) and more importantly, for the polariton ground state (Fig 4b). Once the stimulated scattering exceeds the spin-relaxation in the system, the polarization in the final state can be expected to increase.

This effect can, indeed, occur below a polariton laser transition, where macroscopic populations in the polariton state is only established for short times (the fluctuating regime), and manifests rather as a modified time-averaged polarization than a superlinear increase in output intensity. Hence, the observed notable excess of DOCP for the polariton modes relative to the higher k exciton-polariton states is a clear manifestation of stimulated scattering to the ground state in our system.

**Conclusion**

We have studied the interplay between valley polarization and Zeeman-splitting in a strongly coupled microcavity with an embedded TMDC monolayer. We find that the Valley-Zeeman-Splitting can be directly extracted from the energetically split emission, even for photonic detuning and we find a substantial valley polarization emerging by the spin-relaxation in the reservoirs. Our data can be interpreted in the framework of a previously developed kinetic Boltzmann model, which accounts for the non-linear interplay between mode splitting and circular polarization.

We note that the spinor exciton-polaritons are a fascinating platform to investigate the effect of spins on coherent many-body states, in particular in the condensed regime. Our study clearly paves the way towards such experiments.

**Acknowledgement**

The Würzburg group acknowledges support by the state of Bavaria. C.S. acknowledges support by the European Research Commission (Project unLiMIt-2D). This work has been supported by the Fraunhofer-Gesellschaft zur Förderung der angewandten Forschung e.V. F.E gratefully acknowledge the financial support by the German Federal Ministry of Education and Research via the funding "2D Nanomaterialien für die Nanoskopie der Zukunft". E.S. acknowledges support from the RFBR Grant No. 17-52-10006 and from the Royal Society International Exchange Grant No. IEC/R2/170227. Work of A.K. is supported by Westlake University (Project No. 041020100118). S.T. acknowledges support from NSF DMR-1838443.

**References**

1. Schneider, C. et al. *Two-dimensional semiconductors in the regime of strong light-matter coupling.* Nat. Comm. **9**, 2695 (2018).

2. Flatten, L. C. et al. *Room-temperature exciton-polaritons with two-dimensional $WS_2$.* Sci. Rep. **1** (2016).

3. Lundt, N. et al. *Room temperature Tamm-Plasmon Exciton-Polaritons with a WSe2 monolayer.* Nat. Comm. **7**, 13328 (2016).

4. Liu, X et al. *Strong light–matter coupling in two-dimensional atomic crystals.* Nat. Photon. **9**, 30 (2015).

5. Xiao, D. et al. *Coupled Spin and Valley Physics in Monolayers of $MoS_2$ and Other Group-VI Dichalcogenides.* Phys. Rev. Lett. **108,** 196802 (2012).

6. Xu, X. et al. *Spin and pseudospins in layered transition metal dichalcogenides.* Nat. Phys. **10,** 343 (2014).

7. Mak, K. et al. *Control of valley polarization in monolayer $MoS_2$ by optical helicity.* Nat. Nanotechnol. **7,** 494 (2012).

8. Cao, T. et al. *Valley-selective circular dichroism of monolayer molybdenum disulphide. Nat. Commun.* **3,** 885 (2012).

9. Zhu, B. et al. *Anomalously robust valley polarization and valley coherence in bilayer WS$_2$*. Proc. Natl. Acad. Sci. **111,** 11606 (2014).

10. Jones, A. M. et al. *Optical generation of excitonic valley coherence in monolayer WSe$_2$.* **8,** 6 (2013).

11. Hanbicki, A. T. et al. *Anomalous temperature-dependent spin-valley polarization in monolayer WS$_2$*. Sci. Rep. **6,** 18885 (2016).

12. Seyler, K. L. et al. *Electrical control of second-harmonic generation in a WSe$_2$ monolayer transistor*. Nat. Nanotechnol. **10,** 1 (2015).

13. Lundt, N. *et al. Observation of macroscopic valley-polarized monolayer exciton-polaritons at room temperature.* Phys. Rev. B rapid **96,** 241403 (2017).

14. Dufferwiel, S. *et al. Valley-addressable polaritons in atomically thin semiconductors.* Nat. Photonics **11,** 497 (2017).

15. Sun, Z. *et al. Optical control of room-temperature valley polaritons*. Nat. Photonics **11,** 491 (2017).

16. Chen, Y. et al. *Valley-Polarized Exciton-Polaritons in a Monolayer Semiconductor*. Nat. Photon. **11,** 431 (2017).

17. Lundt, N. *et al. Valley polarized relaxation and upconversion luminescence from Tamm-Plasmon Trion-Polaritons with a MoSe$_2$ monolayer.* 2D Mater. **4,** 25096 (2017).

18. Srivastava, A. *et al. Valley Zeeman effect in elementary optical excitations of monolayer WSe$_2$.* Nat. Phys. **11,** 141 (2015).

19. MacNeill, D. *et al. Breaking of Valley Degeneracy by Magnetic Field in Monolayer.* Phys. Rev. Lett. **114,** 37401 (2015).

20. Li, Y. *et al. Valley Splitting and Polarization by the Zeeman Effect in Monolayer.* Phys. Rev. Lett. **113,** 266804 (2014).

21. Aivazian, G. *et al. Magnetic control of valley pseudospin in monolayer WSe2.* Nat. Phys. **11,** 148 (2015).

22. Scrace, T. *et al. Magnetoluminescence and valley polarized state of a two-dimensional electron gas in WS2 monolayers.* Nat. Nanotechnol. **10,** 603 (2015).

23. Castellanos-Gomez, A. *et al.* Deterministic transfer of two-dimensional materials by all-dry viscoelastic stamping. 2D Mater. **1,** 11002 (2014).

24. Waldherr, M et al. *Observation of bosonic condensation in a hybrid monolayer MoSe2-GaAs microcavity.* Nat. comm. **9**, 3286 (2018).

25. Deng, H. et al. *Exciton-polariton Bose-Einstein condensation*. Rev. Mod. Phys. **82,** 1489 (2010).

26. Values of the fitting parameters are as following: the lifetimes and the spin relaxation times are $\tau = 0.5$ ps, $\tau_R = 500$ ps and $\tau_s = 3.5$ ps, $\tau_{R,s} = 3500$ ps, respectively. The scattering rate is $W = 0.0004$ ps$^{-1}$.